\documentclass[printer]{aa}
\usepackage[varg]{txfonts}
\usepackage{txfonts}
\usepackage{amsmath}

\usepackage[colorlinks]{hyperref}
\hypersetup{
    citecolor = {blue},
    linkcolor = {blue}
}

\begin{document} 
\title{
Outgassing induced acceleration of comet 67P/Churyumov-Gerasimenko
}
\titlerunning{67P outgassing induced acceleration}
\author{T. Kramer \inst{1,2}
\and M. L\"auter \inst{1}
}

\institute{Zuse Institute Berlin, Supercomputing Division, Takustr.\ 7, 14195 Berlin, Germany
\and
Department of Physics, Harvard University, 17 Oxford St, Cambridge, MA 02138, USA
}

\date{Received \today}

\abstract
{
Cometary activity affects the orbital motion and rotation state due to sublimation induced forces.
The availability of precise rotation-axis orientation and position data from the Rosetta mission allows one to accurately determine the outgassing of comet Churyumov-Gerasimenko/67P (67P).
}
{
We derive the observed non-gravitational acceleration of 67P directly from the Rosetta spacecraft trajectory. 
From the non-gravitational acceleration we recover the diurnal outgassing variations and study a possible delay of the sublimation response with respect to the peak solar illumination. 
This allows us to compare the non-gravitational acceleration of 67P with expectations based on empirical models and common assumptions about the sublimation process.
}
{
We use an iterative orbit refinement and Fourier decomposition of the diurnal activity to derive the outgassing induced non-gravitational acceleration.
The uncertainties of the data reduction are established by a sensitivity analysis of an ensemble of best-fit orbits for comet 67P.
}
{
We find that the Marsden non-gravitational acceleration parameters reproduce part of the non-gravitational acceleration but need to be augmented by an analysis of the nucleus geometry and surface illumination to draw conclusions about the sublimation process on the surface.
The non-gravitational acceleration follows closely the subsolar latitude (seasonal illumination), with a small lag angle with respect to local noon around perihelion.
The observed minor changes of the rotation axis do not favor forced precession models for the non-gravitational acceleration.
}
{
In contrast to the sublimation induced torques, the non-gravitational acceleration does not put strong constraints on localized active areas on the nucleus.
We find a close agreement of the orbit deduced non-gravitational acceleration and the water production independently derived from Rosetta in-situ measurement.
}

\maketitle

\section{Introduction}

Studying the non-gravitational acceleration of comets provides important insights into the sublimation of cometary ices.
Earth-bound astrometry allows one to determine the orbital corrections arising from the sublimation of volatiles.
Non-gravitational accelerations are often invoked to explain the orbital evolution of comets and even interstellar objects entering the solar system
(\cite{Whipple1950,Marsden1973,Micheli2018,Sekanina2019}).

As pointed out by \cite{Yeomans2004}
it is therefore of interest to compare widely used models with the precise data returned from spacecraft missions.
For comet 67P \cite{Krolikowska2003} provided an assessment of the Marsden parameters within the asymmetric outgassing model of \cite{Yeomans1989} before the arrival of Rosetta at the comet.
The determination of accurate parameters from Earth bound telescopic observations requires to monitor the position of the comet at several apparitions.

The situation is different for the orbit of 67P as observed by Rosetta.
Rosetta accompanied 67P for more than two years and provided measurements of the three-dimensional position vector with an accuracy better than 100~km (see \cite{Godard2017}).
The discrepancy between Earth bound orbit prediction and the position where Rosetta encountered 67P $590$~days before perihelion requires to adjust the Marsden-type orbit by $2000$~km.
From this it can be estimated that Rosetta provided an up-to twenty times higher accuracy compared to previous orbit determinations.

In conjunction with the precisely known rotation state of 67P we perform an attribution of the actual orbital changes to the sublimation activity across the nucleus.
To account for the observed gas release of 67P requires to extend the non-gravitational acceleration models developed by Whipple, Marsden, and Sekanina.
To this end we introduce a Fourier decomposition of the sublimation induced force and establish a general expression connecting the diurnal variations of the sublimation rate with the orbital evolution.
This formalism simplifies the analysis by eliminating intermediate angles and emphasizes the lack of proportionality between total production rate and non-gravitational acceleration.

\section{Orbit changes by non-gravitational acceleration}

The equation of motion for the position vector $\vec{r}$ of the cometary nucleus is described by contributions of solar system gravitational acceleration $\vec{a}_\mathrm{G}$ and the additional non-gravitational acceleration $\vec{a}_\mathrm{NG}$ (see e.g.\ \cite{Yeomans2004}) 
\begin{equation}\label{eq:orbit}
\ddot{\vec{r}} = \vec{a}_\mathrm{G} + \vec{a}_\mathrm{NG}.
\end{equation}
In general this equation holds in any inertial system but we considered it in the Earth's equatorial system.
The gravitational part $\vec{a}_\mathrm{G} = \vec{a}_\mathrm{G}(\vec{r},t)$ is evaluated at the momentary position vector and takes into account the acceleration due to all solar system planets and the Earth Moon, Pluto, Vesta, and Ceres.
Additional corrections due to relativistic effects are ignored here, since they play only a minor role for 67P.
Relativistic corrections could be added for other objects moving faster around the sun.
The non-gravitational part is affected by the sublimation of volatiles on the nucleus across the surface.
For given initial conditions $\vec{r}(t_0)$, $\dot{\vec{r}}(t_0)$ and a suitable model of the non-gravitational acceleration, the orbit of a comet can be integrated with high precision.

\subsection{Marsden-type orbits}\label{ssec:marsden}

To understand the origin of the non-gravitational acceleration $\vec{a}_\mathrm{NG}$ it is helpful to consider different reference frames.
Starting from the icy snowball model introduced by \cite{Whipple1950,Whipple1951} to account for sublimation processes, Marsden and co-authors developed a powerful parametrization for the non-gravitational acceleration in a series of papers \cite{Marsden1968,Marsden1969,Marsden1970,Marsden1971,Marsden1972,Marsden1973}.
Marsden expressed $\vec{a}_\mathrm{NG}$ with respect to three right-handed orthogonal unit vectors with $\vec{e}_\mathrm{r}$ pointing from the sun to the nucleus, $\vec{e}_\mathrm{n}$ directed along the orbital angular momentum perpendicular to the orbital plane, and $\vec{e}_\mathrm{t}$ being orthogonal to both, $\vec{e}_\mathrm{r}$ and $\vec{e}_\mathrm{n}$.
By comparing 14 cometary orbits \cite{Marsden1973} derived the following model for the non-gravitational acceleration
\begin{equation}\label{eq:marsden}
\vec{a}_{\mathrm{NG},\text{Marsden}}
=g(r') 
(A_1 \vec{e}_r+A_2\vec{e}_t+A_3\vec{e}_n),
\end{equation}
with the constant parameters $A_1$, $A_2$, $A_3$ and the empirical activity function 
\begin{equation}\label{eq:gr}
g(r')=\alpha {\left(\frac{r}{r_0}\right)}^{-m}{\left(1+{\left(\frac{r}{r_0}\right)}^{n}\right)}^{-k}.
\end{equation}
The solar distance $r'=r(t-\Delta t)$ includes a time-shift $\Delta t$ introduced by \cite{Yeomans1989} to account for asymmetries of the activity with respect to the perihelion as studied by \cite{Sekanina1988}.
As we discuss later, $g(r)$ is not directly proportional to the sublimation flux as originally stated by \cite{Marsden1973}, Eq.~(4).
The Marsden parameters provide an excellent, albeit empirical description of the non-gravitational acceleration.
The integration of a Marsden-type orbit proceeds by solving Eq.~\eqref{eq:orbit} with the non-gravitational acceleration Eq.~\eqref{eq:marsden}.

\begin{figure}
\centering
\includegraphics[width=0.99\linewidth]{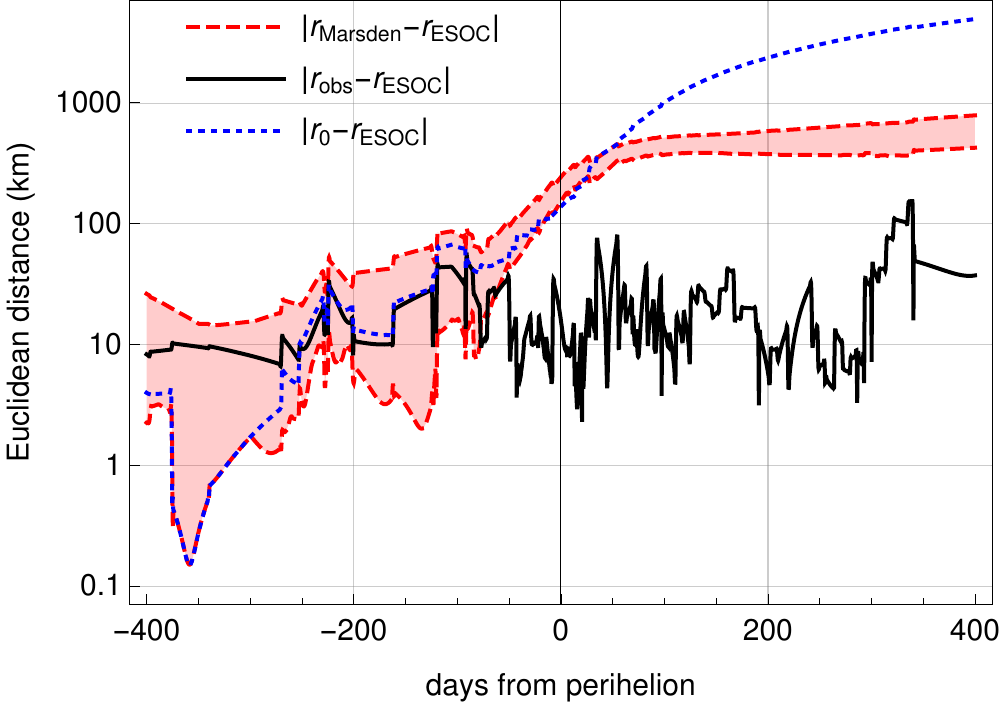}
\includegraphics[width=0.99\linewidth]{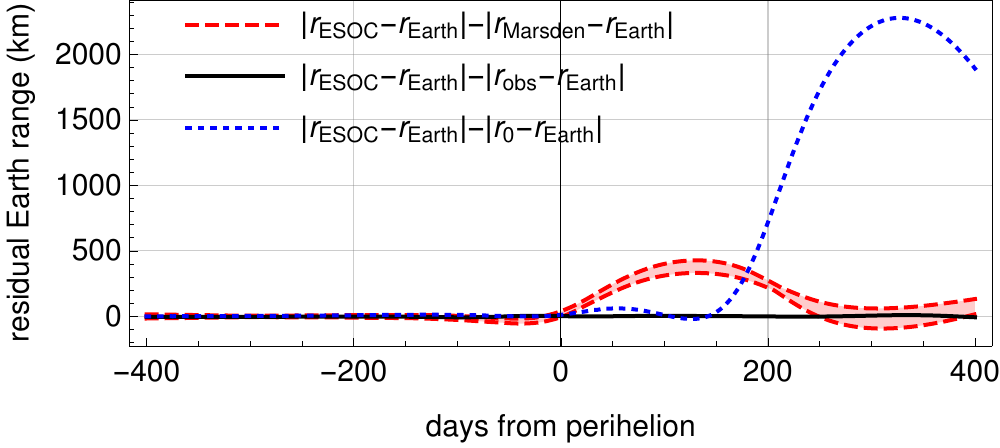}
\caption{
Upper panel: Euclidean error norm of various orbital solutions with respect to the multi-arc ESOC reference.
Lower panel: Residuals with respect to the Earth bound range of the orbital solutions.
 $\mathbf{r}_0$ is the purely gravitational solution, $\mathbf{r}_\text{Marsden}$ a Marsden type orbit, and
$\mathbf{r}_{\rm obs}$ the best fit reconstructed orbit for the non-gravitational acceleration shown in Fig.~\ref{fig:angpqr}.
The shaded band indicates the variation across $31$ initial positions for the Marsden solution.
}
\label{fig:errmarsden}
\end{figure}
\subsection{Orbit determination based on spacecraft data}

To move beyond the Marsden model requires more detailed data from spacecraft missions or radar observations about the magnitude and direction of the observed non-gravitational acceleration in space.
For 67P the Rosetta mission provides a two years data set for both, the rotation-axis orientation (see \cite{Kramer2019}) and the orbital evolution, but required an iterative process to obtain the three-dimensional non-gravitational acceleration with high precision.
We started with the the multi-arc orbital solution from the flight dynamics team at ESOC, which is available as SPICE kernel (CORB\_DV\_257\_03\_\_\_T19\_00345.BSP) and is returned from the Horizons system as position for 67P during the Rosetta mission.
In the following we refer to it as $\vec{r}_\text{ESOC}(t)$.
Before and after the Rosetta mission, Horizons returns the solution 67P/K154/2 based on Marsden parameters with a discontinuous jump by $2000$~km into the Rosetta period.
Besides this discontinuity in Horizons, further discontinuities in the ESOC position vector $\vec{r}_\text{ESOC}(t)$ during the Rosetta mission exist, which do not allow one to obtain the acceleration by a second derivative of the position vector (see the discussion by \cite{Attree2019}).

An accurate estimate of the non-gravitational acceleration is tied to finding the best possible initial condition during a time with negligible cometary activity.
We performed an iterative orbit refinement to identify the initial conditions which minimize the error norm with respect to the ESOC data in the time period $(-400,-200)$~days from perihelion (see Table~\ref{tab:ICorbit}).
During this initial search only the gravitational acceleration was considered, thus we evaluated the cometary orbits with Eq.~\eqref{eq:orbit} with the setup $\vec{a}_\mathrm{NG}=0$.

In the next step we added Marsden-type non-gravitational accelerations with the parameters used by Horizons (see Table~\ref{tab:A123Horizons} and the parameter estimation by \cite{Krolikowska2003}) and we solved  Eqs.~\eqref{eq:orbit}, \eqref{eq:marsden} to obtain $\vec{r}_\text{Marsden}(t)$.
We extended the integration to the years 1959-2022 (limited by two Jupiter encounters) and verified that our initial conditions lead to orbital solutions within $4400$~km with respect to the Horizons solution.
We compared our results with the ones obtained by submitting the corresponding osculating elements to the Advanced Horizons Asteroid \& Comet SPK web interface \cite{Horizons2019}.
For the given initial conditions our Marsden-type solution (Fig.~\ref{fig:errmarsden}) is considerably closer to $\vec{r}_\text{ESOC}$ than the Marsden solution given by \cite{Attree2019}.
\cite{Attree2019} reported a Marsden solution with a difference in Earth bound range
$|\vec{r}_\text{ESOC}-\vec{r}_\text{Earth}|-|\vec{r}_\text{Marsden}-\vec{r}_\text{Earth}|$ of $1500$~km $400$~days after perihelion, while our Marsden solution at that time differs only by $50$~km.
This highlights the necessity to perform a throughout statistical ensemble analysis of initial conditions to validate the non-gravitational part of the orbital acceleration.
In the final iteration we fitted the remaining difference vector $\Delta \vec{r}(t)=\vec{r}_\text{ESOC}-\vec{r}_\text{Marsden}$ to a combination of exponentials and polynomials up to fourth order, which provided (upon adding it to the Marsden solution) a differentiable representation of the observed orbit 
\begin{equation*}
\vec{r}_{\rm obs}(t)=\vec{r}_\text{Marsden}(t)+\Delta \vec{r}_\text{fit}(t)
\end{equation*}
(see Fig.~\ref{fig:errmarsden}).
Only after these iterative steps the non-gravitational acceleration was obtained from
\begin{equation}\label{eq:ngobs}
\vec{a}_{\mathrm{NG},\mathrm{obs}}=\ddot{\mathbf{r}}_\mathrm{obs}(t)
-\vec{a}_{\mathrm{G}}(\mathbf{r}_\mathrm{obs}(t)).
\end{equation}
The validation of the second order derivative of the initially noisy position vector required a careful analysis of errors introduced by the fit.
We have repeated the entire analysis by applying the smoothing and differentiation filter introduced by \cite{Savitzky1964} with the corrections by \cite{Steinier1972} to the larger difference vector $\vec{r}_\text{ESOC}-\vec{r}_0$, where $\vec{r}_0$ includes only gravitational forces.
The ten times larger distance increases the fit uncertainties but leads to qualitative similar results in the period $\pm 100$~days from perihelion.
Remaining systematic errors are discussed in Sect.~\ref{sec:physprop}.
\begin{figure}
\centering
\includegraphics[width=0.99\linewidth]{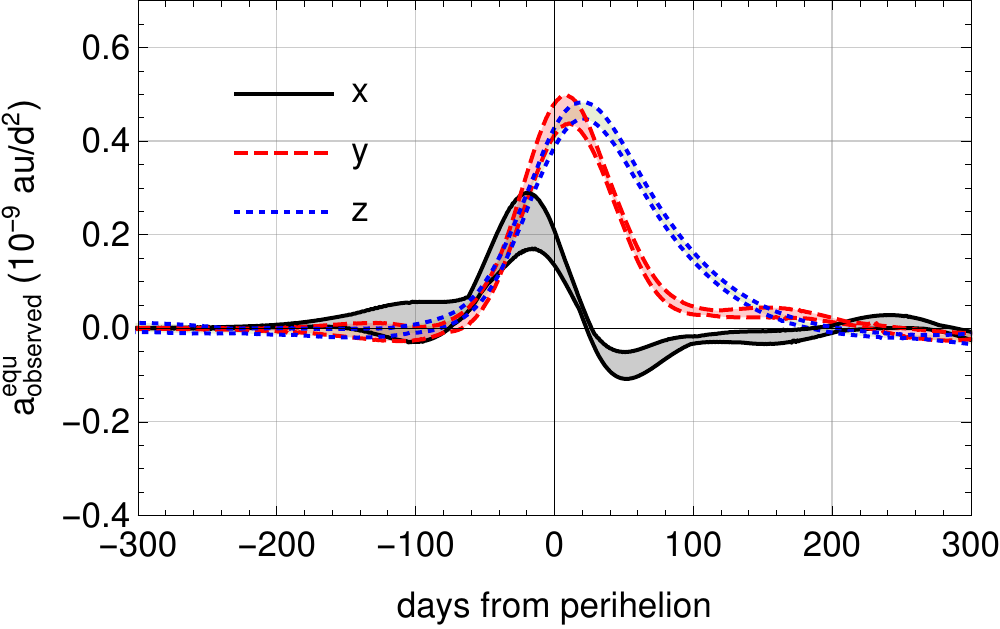}\\
\includegraphics[width=0.99\linewidth]{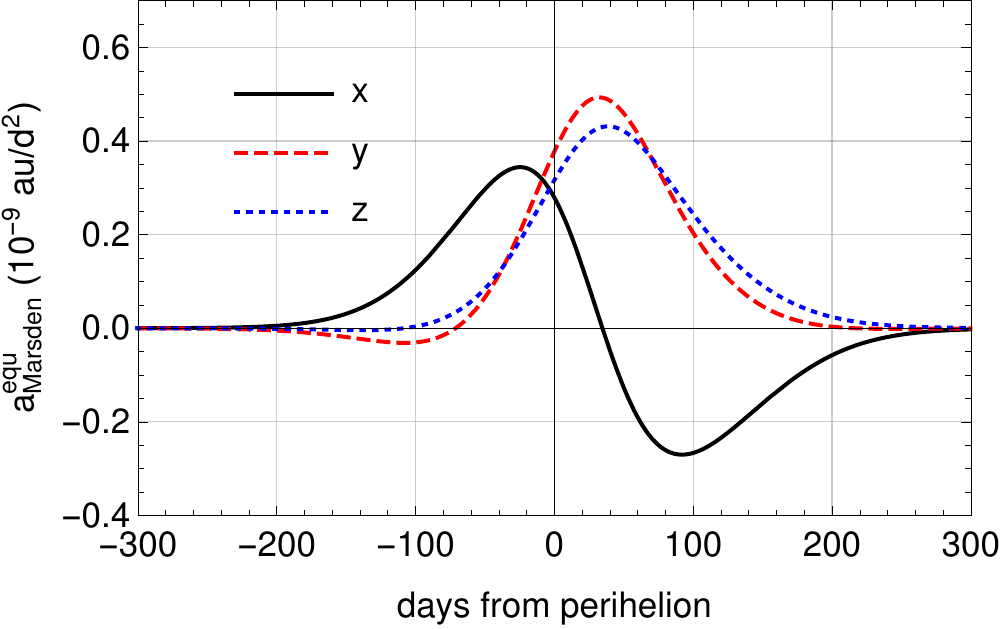}
\caption{Non-gravitational acceleration of 67P in the Earth equatorial frame.
Upper panel: Observed $\vec{a}_{\mathrm{NG},\mathrm{obs}}$. The shaded band indicates the variation across the set of initial conditions. 
Lower panel: Marsden model $\vec{a}_{\mathrm{NG},\mathrm{Marsden}}$. The parameters are from Tables~\ref{tab:ICorbit}, \ref{tab:A123Horizons}.
}
\label{fig:angpqr}
\end{figure}
The resulting non-gravitational acceleration is shown in Fig.~\ref{fig:angpqr}.
In combination with the initial condition (Table~\ref{tab:ICorbit}), the retrieved $\vec{a}_{\mathrm{NG},\mathrm{obs}}$ reproduced the multi-arc solution $\vec{r}_\text{ESOC}$ for $\pm400$ days around perihelion with an mean error of $21$~km, mainly caused by the nonphysical jumps (see black curve in Fig.~\ref{fig:errmarsden}).
At most times the error is around 10~km ($\approx 5$ cometary radii).
To test the sensitivity of the orbit with respect to the initial conditions, we investigated an ensemble of 1000 nearby initial conditions and verified that no further improvement is obtained.
Out of the 1000 initial conditions we identified $31$ orbits with a mean error $<22$~km in the $\pm400$~days interval around perihelion, which all originate from a phase space volume extending about $3$~km around the initial position listed in Table~\ref{tab:ICorbit} and with $<10^{-5}$ variations in the velocities.
In the following we show results for this set of $31$ solutions in the form of shaded bands to estimate the uncertainties of derived quantities from the orbits. 
\begin{table}
\begin{tabular}{ccl}
quantity & value & unit \\\hline
$x$       & $+1.332184491942$  & au \\
$y$       & $-2.770636585665$ & au \\
$z$       & $-1.613532002605$ & au \\
$\dot{x}$ & $0.0042333005604361959$ & au/d \\
$\dot{y}$ & $0.0074132927293021402$ & au/d \\
$\dot{z}$ & $0.0034828287065240359$ & au/d 
\end{tabular}
\caption{Representative initial conditions for the orbit integration at JDB $2456897.71970$ ($\approx-350$~days from perihelion), mean equator and equinox of the Earth J2000 frame.
Additional $30$ nearby initial conditions have been identified which differ on average less than $22$~km over the $(-400,400)$~days interval from the partly nonphysical ESOC orbit. 
}
\label{tab:ICorbit}
\end{table}

\begin{table}
\begin{tabular}{ccl}
quantity & value & unit \\\hline
$A_1$ & $1.066669896245\times 10^{-9}$ & au/d$^2$ \\
$A_2$ & $-3.689152188599\times 10^{-11}$ & au/d$^2$\\
$A_3$ & $2.483436092734\times 10^{-10}$ & au/d$^2$ \\
$\Delta t$ & $35.07142$  & d
\end{tabular}
\caption{Non-gravitational acceleration parameters used by Horizons in conjunction with the standard parameters
$\alpha=0.1112620426$, $k=4.6142$, $m=2.15$, $n=5.093$, $r_0=2.808$ for $g(r)$ in Eq.~(\ref{eq:gr}).
}
\label{tab:A123Horizons}
\end{table}

\section{Transformation of the non-gravitational acceleration to the cometary body}\label{sec:trafo}

Next we connected the observed non-gravitational acceleration to the cometary activity on the surface.
We assume that the nucleus is not in tumbling rotation and that the rotation period and axis orientation are fixed during a single cometary rotation.
For 67P this is an excellent approximation since the orientation over $800$~days changed only by $0.5^\circ$ and the comet rotation period $T_{\rm rot}$ decreased by $21$~minutes from $12.4$~h $300$~days before the 2015 perihelion (see \cite{Godard2017,Kramer2019}).

The Marsden non-gravitational acceleration given by Eq.~\eqref{eq:marsden} is restricted to the direction dictated by the time-independent linear combination $A_1,A_2,A_3$ of the co-moving basis.
This rigid link ignores the physical properties of the nucleus, in particular the rotation axis orientation and rotation period encoded in the angular velocity vector $\vec{\omega}$.
For cometary activity driven by the solar illumination on the nucleus the 
$A_1$, $A_2$, $A_3$ components are no longer time independent.
\cite{Sekanina1967} studied the arising temporal variation of the Marsden parameters under the assumption of a fixed orientation of the rotation axis.
Sekanina introduced a coordinate system that takes into account the obliquity of the comet equatorial plane with the orbital plane to study the illumination conditions of the subsolar point during the orbital motion.
This approach was further extended by \cite{Whipple1979}, \cite{Sekanina1984} and \cite{Sitarski1990} to time-dependent Marsden parameters, including precession models with a changing rotation axis and associated oblateness of the nucleus.
Before the rotation state of 67P was known, \cite{Krolikowska2003} applied different models to 67P including forced precession solutions.

For 67P a detailed shape model is available from \cite{Preusker2017} and the changes of the rotation state are known (see \cite{Jorda2016,Kramer2019}).
67P showed a very repetitive diurnal pattern of gas and dust in the coma across the entire illuminated nucleus, indicating a very regular and periodically repeating outgassing \cite{Kramer2016,Kramer2017,Kramer2018,Lauter2019}.
Of particular interest is the surface activity with respect to the subsolar point in the body frame.
For a given rotation vector $\vec{\omega}$ and position vector $\vec{r}$ of the nucleus the rotation matrix $\tens{R}_{\mathrm{com\rightarrow equ}}$ is the transformation from the cometary equatorial frame (without nucleus rotation)
to the Earth's equatorial frame,
\begin{equation}\label{eq:comtrans}
\tens{R}_{\mathrm{com\rightarrow equ}}
=
\left(
\frac{\vec{h}}{|\vec{h}|},
-\frac{\vec{\omega}\times\vec{r}}{|\vec{\omega}\times\vec{r}|},
\frac{\vec{\omega}}{|\vec{\omega}|}
\right),
\end{equation}
with $\vec{h} = -(\vec{\omega}\times\vec{r})\times\vec{\omega}$.
It puts the sun at a fixed subsolar longitude.
This construction is very similar to Sekanina's system (both share the basis vector $\vec{\omega}/|\vec{\omega}|$).
Any arising force (observed or modeled) from cometary activity $\vec{F}^\mathrm{equ} = \tens{R}_{\mathrm{com\rightarrow equ}} R_{\rm z}(-\omega t)\vec{F}^{\rm bf}$ is expanded in a Fourier series with respect to the subsolar longitude
\begin{equation}\label{eq:FBF}
\vec{F}^{\rm bf}(\lambda_\sun)=
\begin{pmatrix}
D_{x0}-D_{x1} \sin\lambda_\sun-D_{x2}\cos\lambda_\sun+\ldots\\
D_{y0}-D_{y1} \sin\lambda_\sun-D_{y2}\cos\lambda_\sun+\ldots\\
D_{z0}-D_{z1} \sin\lambda_\sun-D_{z2}\cos\lambda_\sun+\ldots
\end{pmatrix}
.
\end{equation}
The Fourier coefficients $D$, in principle defined for each single rotation, are slowly varying functions with the orbital positions around the sun.
This expression encompasses comets with few active regions (see \cite{Jewitt1997} for a simple model of a rectangular shaped comet) as well as globally active ones.
The Fourier representation facilitates the rotational averaging across one rotation period ($\lambda_\sun(t)=-\omega t$ for 67P) in the inertial system
\begin{equation}\label{eq:Fav}
\langle\vec{F}^\mathrm{equ}\rangle
=
\tens{R}_{\mathrm{com\rightarrow equ}} 
\int_0^{T_{\rm rot}}
\frac{{\rm d}t}{T_{\rm rot}}
\tens{R}_{\rm z}(-\omega t)
\vec{F}^{\rm bf}(-\omega t)
\end{equation}
with the rotation matrix around the $z$ axis $\tens{R}_\mathrm{z}$ in the notation of \cite{Montenbruck2000}.
The final expression for the non-gravitational acceleration acting on the orbit of a comet with mass $M$ ($10^{13}$~kg for 67P) is given by inserting Eq.~(\ref{eq:FBF}) into Eq.~(\ref{eq:Fav})
\begin{equation}\label{eq:FF}
\vec{a}_{\mathrm{NG},\mathrm{obs}}=\frac{\langle\vec{F}^{\rm equ}\rangle}{M}
=
\frac{1}{M}
\tens{R}_{\mathrm{com\rightarrow equ}}
\begin{pmatrix}
D_1 \\
D_2 \\
D_3 
\end{pmatrix}
\end{equation}
with the three linear combinations remaining from the complete Fourier expansion
\begin{eqnarray}
\begin{pmatrix}
D_1 \\
D_2 \\
D_3 
\end{pmatrix}
&=&
\begin{pmatrix}
 (-D_{y1}-D_{x2})/2 \\
 (+D_{x1}-D_{y2})/2 \\
 D_{z0} 
\end{pmatrix}\\\nonumber
&=&\sqrt{D_1^2+D_2^2+D_3^2}
\begin{pmatrix}
\cos \phi_D \cos \lambda_D \\
\cos \phi_D \sin \lambda_D \\
\sin \phi_D 
\end{pmatrix}
.
\end{eqnarray}
The parameters $D_1$, $D_2$, $D_3$ are the force coefficients with respect to the cometary equatorial frame (represented by the transformation in Eq.~\eqref{eq:comtrans}).
The link from Eq.~\eqref{eq:comtrans} to the commonly used Sekanina angles is established by expressing 
\begin{equation}\label{eq:commont}
\tens{R}_{\mathrm{com}\rightarrow\mathrm{equ}}
= (\vec{P},\vec{Q},\vec{R}) \tens{R}_\mathrm{z}(\Phi) \tens{R}_\mathrm{x}(-I)
\tens{R}_\mathrm{z}(\pi-\theta_0)
\end{equation}
in terms of the orbital vectors $\vec{P}$, $\vec{Q}$, $\vec{R}$, the obliquity $I$ of the orbit plane to the equator of the comet, the argument of the subsolar meridian at perihelion $\Phi$, and the longitude of the subsolar meridian from the ascending node of the orbit plane on the equator $\theta_0$.
The Marsden basis vectors in Sect.~\ref{ssec:marsden} have the representation
\begin{equation}\label{eq:marmont}
(\vec{e}_\mathrm{r}, \vec{e}_\mathrm{t}, \vec{e}_\mathrm{n}) = 
(\vec{P},\vec{Q},\vec{R}) \tens{R}_\mathrm{z}(-\nu)
\end{equation}
with the true anomaly $\nu$.
The vector with longitude $\lambda_D$ and latitude $\phi_D$ in the cometary equatorial frame transforms to the components $U_\mathrm{r}, U_\mathrm{t}, U_\mathrm{n}$ for the Marsden basis by the relation
\begin{equation}
\begin{pmatrix}
U_\mathrm{r} \\ U_\mathrm{t} \\ U_\mathrm{n}
\end{pmatrix}
=
(\vec{e}_\mathrm{r}, \vec{e}_\mathrm{t}, \vec{e}_\mathrm{n})^{-1}
\tens{R}_{\mathrm{com}\rightarrow\mathrm{equ}}
\begin{pmatrix}
 \cos\phi_D \cos\lambda_D \\ \cos\phi_D \sin\lambda_D \\ \sin \phi_D
\end{pmatrix}.
\end{equation}
Setting $\theta + \pi = \lambda_D$ and $\phi = - \phi_D$ yields Eq.~(2) in \cite{Sekanina1981}.
Assuming additionally $-\phi_D$ to be the latitude of the subsolar point this relation simplifies to the classical Eq.~(4) in \cite{Sekanina1981}.

\section{Physical properties from non-gravitational acceleration}\label{sec:physprop}

\begin{figure}
\centering
\includegraphics[width=0.99\linewidth]{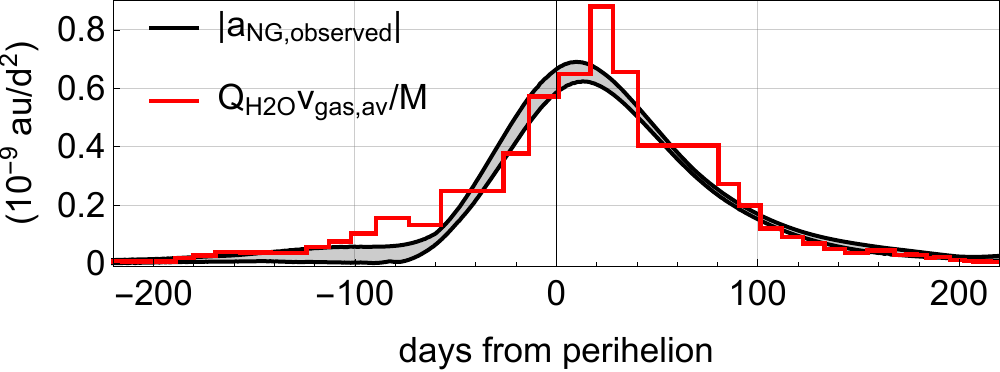}
\includegraphics[width=0.9\linewidth]{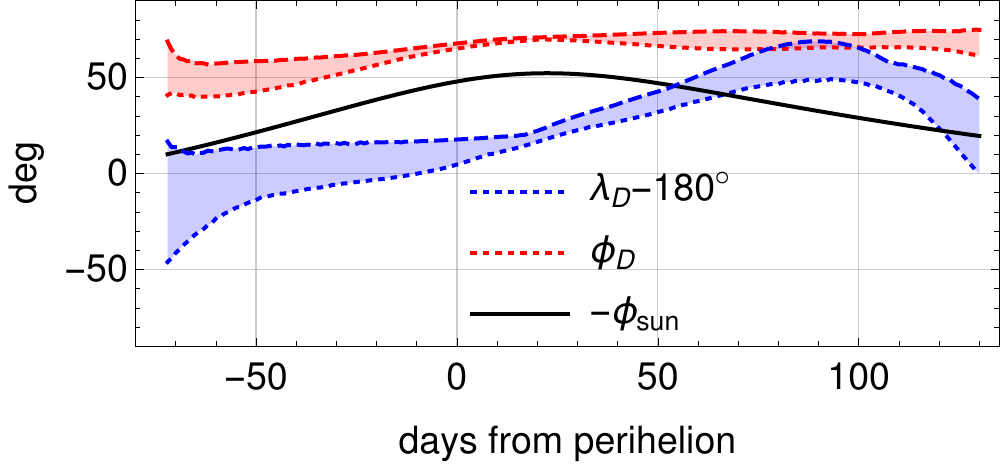}
\caption{
Observed non-gravitational acceleration in the cometary equatorial frame, magnitude (upper panel) and corresponding latitude and longitude of the direction of the sublimation force (lower panel).
The latitude $\phi_D$ is correlated with the anti-solar latitude ($-\phi_{\rm sun}$) (black line in the lower panel), while the longitude correlates with the night terminator $\lambda_D=180^\circ$.
The shaded band indicates the variation across the set of initial conditions.
The red graph in the upper panel shows the ROSINA derived water production of 67P from \cite{Lauter2019}.
} 
\label{fig:drobust}
\end{figure}

The parameters $D_1$, $D_2$, $D_3$ and with it the non-gravitational acceleration arise from the diurnally averaged activity along the spin axis and the amplitudes in the equatorial plane of the comet.
This is reflected in the observed data once decomposed in the cometary equatorial frame in terms of spherical coordinates, see Fig.~\ref{fig:drobust}.
The longitude $\lambda_D$ and latitude $\phi_D$ denote the direction of the acceleration with respect to the subsolar point.
Around perihelion the observed acceleration points towards the north in accordance with the subsolar latitude $\phi_{\rm sun}$ around $-50^\circ$, while at the two equinox crossings the acceleration lies in the equatorial plane.
Up to a shift towards more southern latitudes the seasonal variation of the subsolar latitude is reflected in the observations.
This provides a confirmation of the validity of Sekanina's approach (see Sect.~\ref{sec:trafo}) with respect to the seasonal component.
The determination of the diurnal lag angle with respect to the solar illumination shows larger uncertainties with no lag discernible up to perihelion. 
After perihelion the acceleration vector lags behind (in time) with respect to the momentary anti-solar direction up to $50^\circ$.
This lag disappears around 140 days after perihelion, when the coma gets more and more CO$_2$ dominated (see \cite{Lauter2019}).
The lag cannot be explained by a forced precession model, since the rotation state changes of 67P are small.
In principle a varying surface activity across the nucleus can lead to a lag angle if local surface normal directions do not point towards the sun (see \cite{Samarasinha1996} for an illustration).
\cite{Davidsson2004} studied the variations of the non-gravitational force with respect to varying activity patterns across an ellipsoidal nucleus (see their Fig.~6).
They find small directional changes of the non-gravitational acceleration.
Coma observations of 67P indicate a very repetitive gas and dust release across the entire illuminated surface, see \cite{Kramer2018}.
This is inline with the observation that sublimation models of 67P are not capable of reproducing the reported orbit $\vec{r}_{\mathrm{ESOC}}$ within an Earth-bound range error $<10$~km.
For instance, the model of \cite{Attree2019} results in a deviation of $140$~km with respect to the observed Earth bound range $150$~days after perihelion.
Besides the shape of the nucleus, another possible cause of an diurnal lag angle is a delay between the maximum illumination and the highest gas release in some areas.

From the observation of sunset jets on 67P \cite{Shi2016} obtain thermal delays of peak surface and sublimation temperatures of about $1-2$~h (corresponding to a rotation of $29^\circ-58^\circ$) in the Ma'at region in April 2015.
Their calculations are based on a thermal inertia of $50$~Wm$^{-2}$K$^{-1}$s$^{1/2}$.
At perihelion and at other locations on the surface, the thermal delays might differ from these values, or effects can cancel.
Finally, the observed lag angle could be affected by an (unknown) systematic error of the ESOC orbit, which cannot be detected by the present analysis.
A conclusive attribution of the observed lag angle to a physical process requires further modelling.
\\
The initial motivation for the study of non-gravitational acceleration was to deduce the total production flux of a comet.
The total sublimation flux $Q(r)$ is approximately given by the product of the average gas velocity $v_{\rm gas,av}$ with the magnitude $a_{\mathrm{NG}}$ of the non-gravitational acceleration divided by the cometary mass $M$
\begin{equation}\label{eq:aq}
a_{\mathrm{NG},\mathrm{Marsden}}=\sqrt{A_1^2+A_2^2+A_3^2}\; g(r)\approx Q(r) v_{\rm gas,av}/M
\end{equation}
The Fourier decomposition of the total force allows one to clarify the relation between sublimation flux and non-gravitational acceleration after factoring out the average gas velocity $v_{\rm av}$:
\begin{eqnarray}\label{eq:Mav1}
Q&=&
\int_0^{T_{\rm rot}}
\frac{{\rm d}t}{T_{\rm rot}} 
\int_{\text{surf}}\!\!\!{\rm d}S\; \frac{|\vec{F}(\vec{r}_{\rm bf},t)|}{|\vec{v}_{\rm gas}(\vec{r}_{\rm bf},t)|} \\\nonumber
&\approx&
\int_0^{T_{\rm rot}}
\frac{{\rm d}t}{T_{\rm rot}} 
\left(\int_{\text{surf}}\!\!\!\!\!{\rm d}S\; |\vec{F}(\vec{r}_{\rm bf},t)|\right)
{\left(\int_{\text{surf}}\!\!\!\!\!{\rm d}S\; |\vec{v}_{\rm gas}(\vec{r}_{\rm bf},t)|\right)}^{-1}.
\end{eqnarray}
Next, we applied the triangle inequality to obtain a lower bound for the sublimation flux
\begin{equation}
Q \ge \int_0^{T_{\rm rot}}
\!\!\!\!\!
\frac{{\rm d}t}{T_{\rm rot}\; v_{\rm gas,av}} \left| \vec{F}^{\rm bf}(-\omega t)\right|,\;
v_{\rm gas,av}=\int_{\text{surf}}\!\!\!\!\!{\rm d}S\; |\vec{v}_{\rm gas}(\vec{r}_{\rm bf},t)|
.
\end{equation}
Inserting Eq.~(\ref{eq:FBF}) for the force components and neglecting faster oscillating terms allowed us to perform the integration analytically 
\begin{equation}\label{eq:Mav}
Q\approx\frac{4 \sqrt{\Delta}}{2\pi v_{\rm gas,av}}
{\rm E}\left( \frac{-D_{x1}^2-D_{y1}^2-D_{z1}^2+D_{x2}^2+D_{y2}^2+D_{z2}^2}{\Delta}  \right),
\end{equation}
where $\Delta=D_{x0}^2+D_{y0}^2+D_{z0}^2+D_{x2}^2+D_{y2}^2+D_{z2}^2$, ${\rm E}(x)$ denotes the complete elliptic integral.
The result shows that the magnitude of the non-gravitational acceleration contains additional force components  besides the $D_1$, $D_2$, $D_3$ components, which lead to minor corrections.

An entirely independent determination of the gas production for 67P has been performed by \cite{Lauter2019} by analyzing the in-situ data of the Double Focusing Mass Spectrometer (DFMS) of the Rosetta Orbiter Spectrometer for Ion and Neutral Analysis (ROSINA).
The non-gravitational acceleration by Eqs.~(\ref{eq:aq}) shows a remarkable agreement to the ROSINA derived water production $Q_{\rm H2O}(r)$ of 67P from \cite{Lauter2019}, Fig.~\ref{fig:drobust}, with gas velocity $v_{\rm gas,av}=480$~m/s.

\section{Conclusions}

The Rosetta mission to comet 67P provided the unique opportunity to retrieve the non-gravitational acceleration of a comet independently from Earth bound observations.
In conjunction with the known rotation state the position data derived from Rosetta allowed us to verify commonly invoked assumptions about the non-gravitational acceleration.
We have shown the sensitivity of the orbit reconstruction to the initial conditions and identified a phase-space volume 350 days before perihelion that leads to orbital solutions following the reported ESOC positions within a mean deviation of $22$~km.
This close match allowed us to extract the three-dimensional non-gravitational acceleration and to relate it to the activity on the nucleus.

Using a Fourier series we have decomposed the non-gravitational acceleration into the averaged outgassing along the rotation axis and the amplitudes of the outgassing along the equatorial plane of the comet.
A similar analysis and Fourier theory has been carried out by \cite{Kramer2019} for the rotation state of 67P.
We provided error bounds for all derived quantities based on an extensive analysis of initial conditions for the orbital integration.
We find up to perihelion no clear signal of a lag angle between illumination and force direction, while at later times deviations from the instantaneous illumination become apparent.
The seasonal effect of the solar illumination on the non-gravitational acceleration is reflected by a strong correlation of the subsolar latitude and the sublimation force direction.
The agreement of the observed non-gravitational acceleration with the seasonal illumination conditions demonstrates that the non-gravitational acceleration can be explained by the water-ice distribution on the entire nucleus.
The derivation of the diurnal lag carries considerably larger errors at times later than 100 days after perihelion, but points to a shift toward larger delay times 50-100 days after perihelion.
The non-gravitational acceleration alone is most indicative about active areas in terms of a zonal mean.

Due to the lack of longitudinal variations the detection of local active areas across the surface can not be expected.
This is in contrast to the torque affecting the rotation axis orientation and rotation period: both react sensitively to local activity variations (\cite{Kramer2019}).

Finally, we have shown that solely based on the analysis of the Rosetta orbit a close agreement with the in-situ measurements of the gas coma and production of 67P exists.
This connection allows one to relate Earth bound astrometry and production estimates with accurate in-situ measurements and models of cometary activity, a prerequisite to advance non-gravitational acceleration models.

\begin{acknowledgements}
The authors acknowledge the North-German Supercomputing Alliance (HLRN) for providing computing time on the Cray XC40.
We thank the anonymous referee for helpful comments.
\end{acknowledgements}

\end{document}